**CLIMATE-SENSITIVE SPECIES**

**Ecology of the cold-adapted species *Nebria germari* (Coleoptera: Carabidae): the role of supraglacial stony debris as refugium during the current interglacial period.**

BARBARA VALLE*[1], ROBERTO AMBROSINI[2], MARCO CACCIANIGA[3], MAURO GOBBI[4]

[1] Department of Biosciences, Università degli Studi di Milano, Via Celoria 26 - 20133 Milano, Italy, e-mail: barbara.valle@unimi.it, https://orcid.org/0000-0003-4829-4776

[2] Department of Environmental science and Policy, Università degli Studi di Milano, Via Celoria 26 - 20133 Milano Milano, Italy, e-mail: roberto.ambrosini@unimi.it, https://orcid.org/0000-0002-7148-1468

[3] Department of Biosciences, Università degli Studi di Milano, Via Celoria 26 - 20133 Milano, Italy. E-mail: marco.caccianiga@unimi.it, https://orcid.org/0000-0001-9715-1830

[4] MUSE-Science Museum of Trento, Corso del Lavoro e della Scienza 3, 38122 Trento (Italy). E-mail: Mauro.Gobbi@muse.it, https://orcid.org/0000-0002-1704-4857

*corresponding author

*In the current scenario of climate change, cold-adapted insects are among the most threatened organisms in high-altitude habitats of the Alps. Upslope shifts and changes in phenology are two of the most investigated responses to climate change, but there is an increasing interest in evaluating the presence of high-altitude landforms acting as refugia.*

*Nebria germari Heer, 1837 (Coleoptera: Carabidae) is a hygrophilic and cold-adapted species that still exhibits large populations on supraglacial debris of the Eastern Alps.*

*This work aims at describing the ecology and phenology of the populations living on supraglacial debris. To this end, we analysed the populations from three Dolomitic glaciers whose surfaces are partially covered by stony debris.*

*We found that supraglacial debris is characterised by more stable colder and wetter conditions than the surrounding debris slopes and by a shorter snow-free period. The populations found on supraglacial debris were spring breeders, differently from those documented in the 1980s on Dolomitic high alpine grasslands, which were reported as autumn breeders. Currently Nebria germari seems therefore to find a suitable habitat on supraglacial debris, where micrometeorological conditions are appropriate for its life-cycle and competition and predation are reduced.*

# INTRODUCTION

In the current scenario of global warming, high altitude habitats are among the most threatened (CAUVY-FRAUNIÉ & DANGLES 2019; FATTORINI *et al* 2020). Some of the most visible effects of climate change are the reduction in glaciers mass and surface and the increase of debris coverage on their surface, because of the reduction of the pressure of the ice volume on the glacier headwalls and the amplification of frost and heat weathering that increases erosion. This phenomenon is transforming many Alpine glaciers into debris-covered glaciers (TIELIDZE *et al* 2020, KRAAIJENBRINK *et al* 2017, CITTERIO *et al* 2007, PAUL *et al* 2007). The debris cover has a strong impact on the dynamic of glaciers: when the debris is thicker than 3-5 cm, it reduces the rate of ice ablation, with critical thickness depending on the lithological nature of the debris and the climatic regime (NAKAWO & RANA 1999).

Under the current climate change scenario, debris-covered glaciers are assuming a key biological role in high-altitude environments because they are able to host cold-adapted arthropod and plant species (HÅGVAR *et al* 2020; CACCIANIGA *et al* 2011; GOBBI *et al* 2011). In fact, cold-adapted species may react to rising temperatures by finding *refugium* areas in glacial and periglacial landforms, as suggested for the Alps by GOBBI *et al* (2011, 2014, 2018). These landforms are defined "cold spots" of biodiversity, because glaciers host few species, which are, however, extremely specialized and exclusive (including many endemics), but currently threatened of extinction (CAUVY-FRAUNIÉ & DANGLES 2019; GOBBI and LENCIONI, 2020). Together with the study of habitats acting as potential refugia for cold adapted species, the knowledge of the ecological needs of climate-sensitive species in these cold habitats is fundamental to understand how climate change will affect Alpine biodiversity.

The ground beetle *Nebria germari* Heer, 1837 (Coleoptera: Carabidae) is a cold-adapted and hygrophilic species (BRANDMAYR & ZETTO BRANDMAYR 1988, GEREBEN 1995, KAUFMANN & JUEN 2001). It can be defined a climate-sensitive species because it is currently restricted to high altitude habitats of the Eastern Alps, from Tessin, Switzerland, to the Prokletije Massif, Albania (LEDOUX & ROUX 2005). This species also exhibits a fragmented distribution pattern (LEDOUX & ROUX 2005), suggesting refugial occurrence, and local scale extinctions were already documented (PIZZOLOTTO *et al* 2014). Moreover, *N. germari* is a brachypterous species with low dispersal ability and nocturnal foraging behaviour. High humidity and low temperatures are known to be the main environmental features affecting its distribution and abundance (PIZZOLOTTO *et al* 2014, KAUFMANN & JUEN 2001, BRANDMAYR & ZETTO BRANDMAYR 1988). It prefers open grounds with low vegetation cover, and substrates with a high percentage of gravel (KAUFMANN & JUEN 2001, BRANDMAYR & ZETTO BRANDMAYR 1988). It lives on scree slopes, along glacier forelands, on rock glaciers and on debris-covered glaciers (KAUFMANN & JUEN 2001; GOBBI *et al* 2014, 2017, GEREBEN *et al* 2011). In these habitats, *N. germari* co-occurs with other ground-dwelling arthropods; the most common on debris-covered glaciers being spiders (Arachida: Araneae), springtails (Hexapoda, Collembola), and less frequently centipedes (Myriapoda: Chilopoda). Springtails are at the base of *Nebria germari* diet (SINT *et al* 2018). Intraguild predation was demonstrated in arthropods living near the glacier fronts (SINT *et al* 2018, RASO *et al* 2014, KÖNIG *et al* 2011), thus, we can hypothesise that the co-existence (*sensu* STAPLES *et al* 2016) with spiders might affect size of *N. germari* populations, particularly in glacier and near-glacier habitats. For instance, spiders may compete with *N. germari* in hunting activity: even if *N. germari* is primarily nocturnal and spiders are mainly diurnal, in this harsh environment *N. germari* activity was documented also during the day (GEREBEN 1995). Other competitors are centipedes, predators that in few cases were found in the same habitat of *N. germari* (unpublished data).

In this study, we aimed at describing the phenology and the current population dynamic of *N. germari* on supraglacial debris in relation to both biotic (competitors and prey availability) and abiotic factors (micrometeorological conditions and soil gravel fraction). In addition, in order to investigate how global warming might affect this species, we compared our data on the present distribution of this species with those found in literature to highlight possible ecological changes that have occurred in the last 60 year. Specifically, we focused on past and present altitudinal distribution of this species, which seems to respond to climate warming (PIZZOLOTTO *et al* 2014), and to its past and current phenology. Indeed, timing of life-history events is crucial for insect species living in high altitude habitats, where the favourable season for the development, growth and reproduction lasts few weeks (SOTA 1996). We expect an upward shift in the distribution, and an earlier phenology of this species in present compared to past studies.

## MATHERIAL AND METHODS

- STUDY AREA

We investigated three populations of *Nebria germari* located on three glaciers of the Dolomites (Italy): Western Sorapiss and Central Sorapiss Glaciers (46°30'43.34''-12°12' 25.12'' and 46°30'43.55'-12°13' 20.75'' Ampezzo Dolomites) and Vedretta d'Agola Glacier (46°9'6.29''-10°51'29.01'' Brenta Dolomites) (Fig 1). In these sites, *N. germari* occurs almost exclusively on supraglacial debris, characterized by a vegetation cover always < 5% and consisting of sparse young individuals of pioneer plant species typical of carbonatic substrata of the alliance Thlaspion rotundifolii Jenny-Lips 1930, with only a very sporadic occurrence on the nearby moraines (BERNASCONI *et al* 2019, VALLE 2019).).

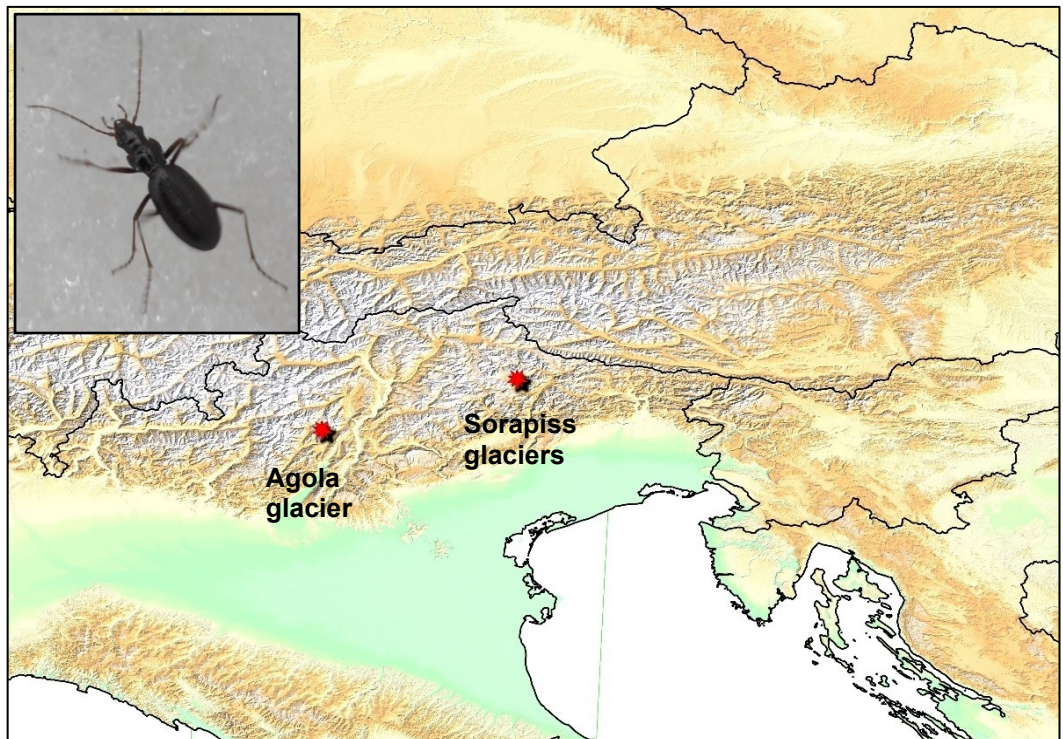

*Figure 1 Map with the position of the two sampling sites (Agola and Sorapiss). At the top left, a picture of N. germari taken on Sorapiss*

All these glaciers of the Dolomites are characterized by relatively low altitude and small surface (Table 1). Climatic features of the investigated areas were extrapolated from closely located weather stations (see micrometeorological analysis paragraph); average annual and monthly temperatures and average annual precipitations are reported in Tables 2 and 3.

During the field work (2017-2018), data on the minimum altitude of glacier systems were collected and the extent of the glaciers and the percentage of debris-covered surface were calculated using aerial photographs from 2015 (Table 1). In addition, debris thickness was estimated with different methods (Table 1). On Western and Central Sorapiss Glaciers debris thickness was measured digging the debris up to the ice along a regular grid of points spaced 20 m to one another (29 points for Wester Sorapiss and 51 for Central Sorapiss, due to different size of the sampling areas). On Agola we measured the thickness around the traps because the use of a regular grid was impossible due to the small extent of the debris-covered part of the Glacier.

| Glacier | Type of glacier | min altitude (m asl) | max altitude (m asl) | Sampling altitude (m asl) | Surface (km$^2$) | % of glacier surface covered by debris | Estimated debris thickness |
|---|---|---|---|---|---|---|---|
| **Agola** | Active mountain glacier | 2601 | 2873 | 2596-2603 | 0.17 | 1.2 | 10-15 cm |
| **Western Sorapiss** | Active mountain glacier | 2254 | 2756 | 2382-2419 | 0.134 | 32 | >50 cm at 50% of measurement points |
| **Central Sorapiss** | Inactive glacier - glacieret | 2213 | 2621 | 2210-2264 | 0.08 | 57.9 | >50 cm at 80% of measurement points |

*Table 1 Altitude and surface of three glacial site. Agola measurements refers to 2018, Sorapiss measurements to 2017. The extent of the glaciers and the degree of debris covered surface were calculated using aerial photographs dated 2015 and field data (2017-20*

- FIELD PROCEDURES

Sampling was carried out during the snow-free period of 2017 (Western and Central Sorapiss Glaciers) and of 2018 (Agola Glacier) by pitfall traps put on the supraglacial debris. Specifically, two plots were selected on Agola, two on Western Sorapiss and four on Central Sorapiss (two on the glacier and two on a dead-ice area separated from the glacier). Each plot was characterised by similar conditions in grain size distribution and vegetation cover (see sampling plan in Appendix 1). Each plot consisted of three pitfall traps - spaced from each other by at least 10 meters. Pitfall traps consisted of a plastic glass (diameter 7 cm) filled up with a non-toxic and frost-resistant solution to catch and preserve the animals during the activity period of the trap. The solution was made by one litre of water, ½ litre of wine-vinegar, one tablespoon of salt and two-three drops of soap. This solution is slightly different from those used in similar studies (e.g. GOBBI *et al* 2017) and was used to reduce its attractive smell toward marmots (Marmota marmota) and alpine ibexes (Capra ibex), which are the most common vertebrates that damage the traps. Three sampling sessions were performed on Agola (06/VIII/18 - 27/VIII/18 - 18/IX/18), four on Western Sorapiss (05/VII/17 - 27/VII/17 - 17/VIII/17 - 07/IX/17), three on Central Sorapiss (28/VII/17 - 18/VIII/17 - 07/IX/17). Sampled specimens of *N. germari* were analysed at the stereomicroscope in order to distinguish them in three developmental stages: larva, teneral (i.e. newly hatched adults) and adult.

A sample of approximately 2 kg of soil was taken from each plot and used to estimate the grain size distributions by dry, wet sieving and aerometry.

Temperature and humidity (micro-meteorological variables hereafter), were recorded every 60 minutes on the supraglacial debris of each glacier by a datalogger (Tinytag plus 2,), positioned 10 cm below the surface close to the plots (see Appendix 1). One datalogger was also placed, on the recent moraine of 1908 (Agola) or 1920 (on Western and Central Sorapiss), respectively 500 and 200 m from the glacier front, to record the same variables on ice-free debris. Datalogger recording periods started from the beginning of the sampling period and were: 17/VI/2018 – 02/VI/2019 on Agola, 22/VI/2017 - 22/VI/2018 on Western Sorapiss, and 07/VII/2017 – 22/VIII/2018 on Central Sorapiss.

- ANALYSIS OF MICROMETEOROLOGICAL DATA

Day and night values were analysed separately because *Nebria germari* is known to be mainly a nocturnal predator (GEREBEN 1995). The average sunrise and sunset times for each month was used to assess day and night measures. The average annual temperature, the average day and the night summer temperature (mean of all daily day and night averages collected from the beginning of sampling period to the end of the snow-free season) were considered. Humidity was not recorded on Central Sorapiss Centrale because the humidity sensor was not available. Although we are aware that conditions can vary according to debris depth and grain size, we are confident that the data we collected are representative of the conditions of the sampling sites.

Supraglacial temperatures of each glacier were compared with values detected on the moraines of 1908/1920 and with atmospheric values calculated on the base of the data from the nearest weather stations with complete available data (Val d'Ambiez weather station for Agola Glacier, 1888 m asl, Trento Province, 3 km apart from the sampling site; Faloria weather station for Sorapiss Glaciers, 1743 m asl, Belluno Province, 4 km apart from the sampling site). In order to correctly compare datalogger and weather station values, which were recorded at different altitudes, we applied to weather station values a correction factor equal to the average vertical thermal gradient for each site and month (Tables 2 e 3). These correction factors were calculated for each site and month by linear interpolation of the data from weather stations located at different altitudes. For Agola Glacier, we used data from Meteotrentino weather stations of Alimonta refuge (2577 m asl, Trento Province), Val d'Ambiez (see above) and Giustino (877 m asl, Trento Province). Available data spanned 2011-2018. For Sorapiss, we used data from ARPA Veneto weather stations of Cortina d'Ampezzo (2335 m asl, Belluno Province) and Faloria (see above). Available data spanned the period 1994-2018.

| Altitude (m asl) | \multicolumn{12}{c}{Mean temperature (°C) in} | Annual average temperature (°C) | Annual average precipitations (mm/y) |
|---|---|---|---|---|---|---|---|---|---|---|---|---|---|---|
|  | J | F | M | A | M | J | J | A | S | O | N | D |  |  |
| 877 | 0.7 | 4.1 | 8.3 | 11.1 | 15.4 | 17.7 | 18.7 | 15.2 | 11.1 | 6.9 | 2.1 | 0.9 | 9.4 | - |
| 1888 | -4.2 | -2.0 | 1.2 | 4.1 | 8.2 | 11.5 | 12.9 | 10.2 | 6.2 | 3.8 | -1.1 | -2.3 | 4.0 | - |
| 2577 | -6.0 | -3.1 | -0.1 | 2.7 | 6.5 | 8.3 | 8.4 | 4.9 | 2.5 | -2.4 | -5.5 | -5.7 | 0.9 | 1000 |
| **Vertical gradient Δt/100m** | -0.20 | -0.21 | -0.25 | -0.26 | -0.26 | -0.28 | -0.30 | -0.30 | -0.25 | -0.27 | -0.22 | -0.19 | -0.25 |  |

*Table 2: monthly vertical thermal gradient for Agola site, with the average data of each weather station for the period 2011-2018 (Alimonta= 2577 m asl, Val d'Ambiez = 1888 m asl, Giustino = 877 m asl) used for the interpolation.*

| Altitude (m asl) | Mean temperature (°C) in | | | | | | | | | | | | Annual average temperature (°C) | Annual average precipitations (mm/y) |
|---|---|---|---|---|---|---|---|---|---|---|---|---|---|---|
| | J | F | M | A | M | J | J | A | S | O | N | D | | |
| 1271 | -1.8 | -0.9 | 2.4 | 5.8 | 10.3 | 13.8 | 15.8 | 15.4 | 11.2 | 7.3 | 2.4 | -1.0 | 8.0 | - |
| 2235 | -5.2 | -5.5 | -3.1 | -0.5 | 4.2 | 8.1 | 10.2 | 10.0 | 6.1 | 3.2 | -1.4 | -4.3 | 1.8 | 1100 |
| Vertical gradient Δt/100m | -0.54 | -0.48 | -0.32 | -0.65 | -0.63 | -0.59 | -0.58 | -0.56 | -0.53 | -0.43 | -0.39 | -0.34 | -0.50 | |

*Table 3: monthly vertical thermal gradient for Sorapiss site, with the average data of each weather station for the period 1994-2018 (Cortina d'Ampezzo = 2235 m asl, Faloria = 1271 m asl) used for the interpolation.*

- ECOLOGICAL AND PHENOLOGICAL DATA ANALYSIS

We considered the following information collected at each trap: days of trap activity, absolute abundance of *N. germari* larvae, subadults and adults, absolute abundance of centipedes, springtails, adult spiders (sum of all Araneae taxa found: *Acantholycosa* spp., families Lyniphiidae and Thomisidae) and gravel percentage. We then used Generalized Linear Models (GLMs) assuming a Poisson data distribution to model the abundance of different *N. germari* developmental stages according to a set of predictors selected in agreement with literature (see introduction), namely: day of the year (1 January = 1), gravel percentage, the abundance of centipedes and springtails (potential prey of *N. germari*), and abundance of spiders (potential predators/competitors of *N. germari*). A three-level factor indicating the glacier was also entered in the models together with the log10-transformed number of days for which a trap was active; the latter variable was entered as an offset to account for difference in sampling effort. With this parameterization, the GLM modelled the average number of individuals collected per day of trap activity. We also accounted for possible data overdispersion, which may inflate type-I error rate of the model, by using the family = quasipoisson option in R. Analyses were performed with R 3.6.2 (R Core team 2019, Foundation for Statistical Computing, Vienna, Austria. URL https://www.R-project.org/.).

# RESULTS

A total of 138 individuals of *Nebria germari* were collected on Agola Glacier (51 larvae, 81 subadults, 6 adults), 257 on Western Sorapiss Glacier (51 larvae, 9 subadults, 197 adults) and 444 on Central Sorapiss Glacier (104 larvae, 82 subadults, 258 adults). The average activity density (AD: number of collected individuals per day of trap activity) of all *N. germari* (larvae, adults and subadults) recorded on each glacier during the whole sampling periods was higher on Sorapiss glaciers (0.64 ± 0.64 on Western Sorapiss and 0.64 ± 0.41 on Central Sorapiss) than on Agola glacier (0.38 ± 0.45).

- TEMPERATURE AND HUMIDITY

We estimated the duration of the snow cover from the daily trend of day and night temperatures (see Appendix2). The beginning of snow cover is indicated by a monotonous decreasing trend of daily temperatures. This occurred on October 28[th] on Agola Glacier and on November 6[th] on Sorapiss Glaciers. The long spring plateau at 0°C indicates that melting was occurring. The end of this plateau indicates the beginning of the snow-free period. On Western Sorapiss snow completely melted by June 30[th], on Central Sorapiss by July 25[th]. The datalogger on Agola Glacier did not record data from 3/VII/2019 to 16/VI/2019. To estimate the beginning of snow-free season at this site, we therefore used data collected by the datalogger placed on the 1908 moraine to estimate the beginning of the snow-free period. On the moraine snow melted on June 30[th]. Thus, on supraglacial debris this happened between 30/VI/2019 and 16/VI/2019 (the end of missing data period). We used these data to calculate thermal conditions linked to our sampling period, reported in Table 4; for Agola Glacier it was not possible due to missing data for June and July.

Day temperature during the snow-free period was almost equal to (on Sorapiss) night temperature (Table 4).

In Figure 2 monthly temperature trends on supraglacial debris, on ice-free debris and in atmosphere, are reported for each glacier (values are reported in Appendix 3). Supraglacial habitat showed colder condition than the atmosphere and the ice-free debris during summer and warmer conditions than the atmosphere during winter. The annual average temperature of supraglacial debris is always lower than that of ice-free debris and atmosphere. The annual temperature is always lower on supraglacial debris. However, each datalogger showed different values, probably because of different thickness and porosity of the supraglacial debris.

| Supraglacial debris of | Average annual temperature (°C) | St. Dev | Average day temperature during snow-free period (°C) | St. Dev | Average night temperature during snow-free period (°C) | St. Dev | Snow-free period (days) |
|---|---|---|---|---|---|---|---|
| Western Sorapiss | **-0.2** | 2.8 | **2.8** | 2.7 | **2.9** | 2.7 | **130** |
| Central Sorapiss | **-0.8** | 1.4 | **1.1** | 1.6 | **0.8** | 1.2 | **105** |

*Table 4 Average temperatures during the year and during the snow-free period and length of the snow-free period at each glacier.*

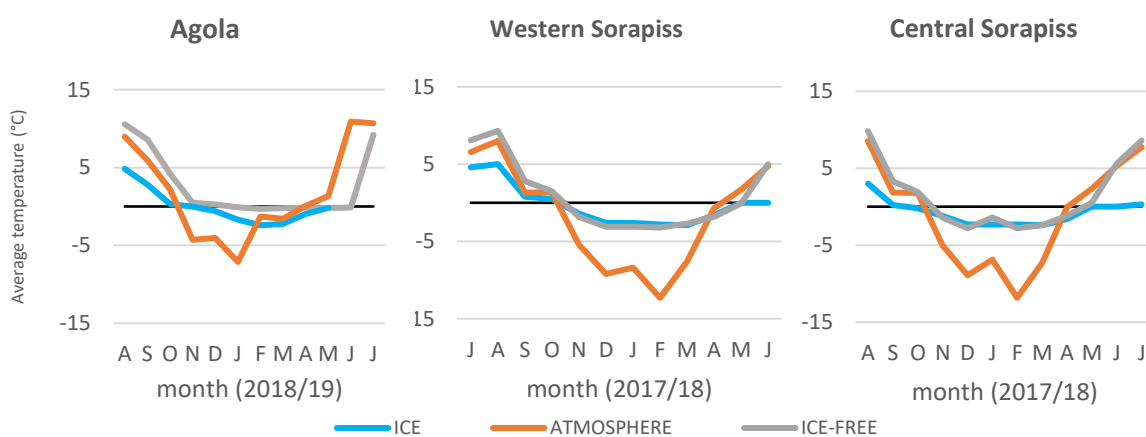

*Figure 2 monthly temperatures on supraglacial debris, ice-free debris and atmosphere. Monthly temperature is the average of daily temperatures per month: light blue indicates supraglacial debris temperatures, orange indicates atmosphere temperatures, grey indicates ice-free debris temperatures*

Maximum recorded values were: 16°C on Agola (23/VII/2018), 10.2°C on Western Sorapiss (07/VII/2017) and 6.9°C Central Sorapiss (20/VII/2017). All daily values of day and night humidity are reported in Appendix 4. Missing data are due to the reset of the sensor when it remains saturated of water for a long time. On the glacier, air under surface showed high levels of relative humidity, with an average daily humidity of 94% with almost no difference between the day and night values.

- ECOLOGICAL AND PHENOLOGICAL DATA ANALYSIS WITH GLM

The abundance of the different developmental stages at each sampling session and glacier is shown in Figure 3. Abundance of both larvae and subadults increased with day-of-year (larvae: coef. = 0.03 ± 0.01 SE, $F_{1,64}$ = 9.10, P = 0.004; subadults: coef. = 0.04 ± 0.01 SE, $F_{1,64}$ = 10.13, P = 0.002), while that of adults decreased along the season (coef. = -0.04 ± 0.01 SE, $F_{1,64}$ = 43.90, P < 0.001).

Abundance of larvae did not differ among glaciers ($F_{2,64}$ = 0.66, P = 0.528), There was a tendency toward differences in overall abundances of subadults among glaciers ($F_{2,64}$ = 3.54, P = 0.035); however, Tukey post-hoc tests failed in detecting any pairwise difference between glaciers (|z| ≤ 2.04, P ≥ 0.095). In contrast, abundance of adults was significantly lower on Amola glacier than on both Sorapiss glaciers ($F_{2,64}$ = 9.62, P < 0.001, post-hoc tests: |z| ≥ 2.71, P ≤ 0.016), which did not differ to one another (z = 0.73, P = 0.728).

Abundance of the different life-stages was also affected by different ecological variables of sampling sites. As shown by the GLM analysis, larvae abundance increased with increasing abundance of chilopods (coef: 0.06 ± 0.02 SE, $F_{1,64}$ = 8.15, P = 0.006), subadult abundance decreased with increasing number of spiders (coef: -0.57 ± 0.27 SE, $F_{1,64}$ = 6.14, P = 0.016), adult abundance decreased at increasing gravel percentages (coef. = -5.10 ± 1.44 SE, $F_{1,64}$ = 13.22, P < 0.001. All the other effects were not significant ($F_{1,64}$ ≤ 2.54, P ≥ 0.115).

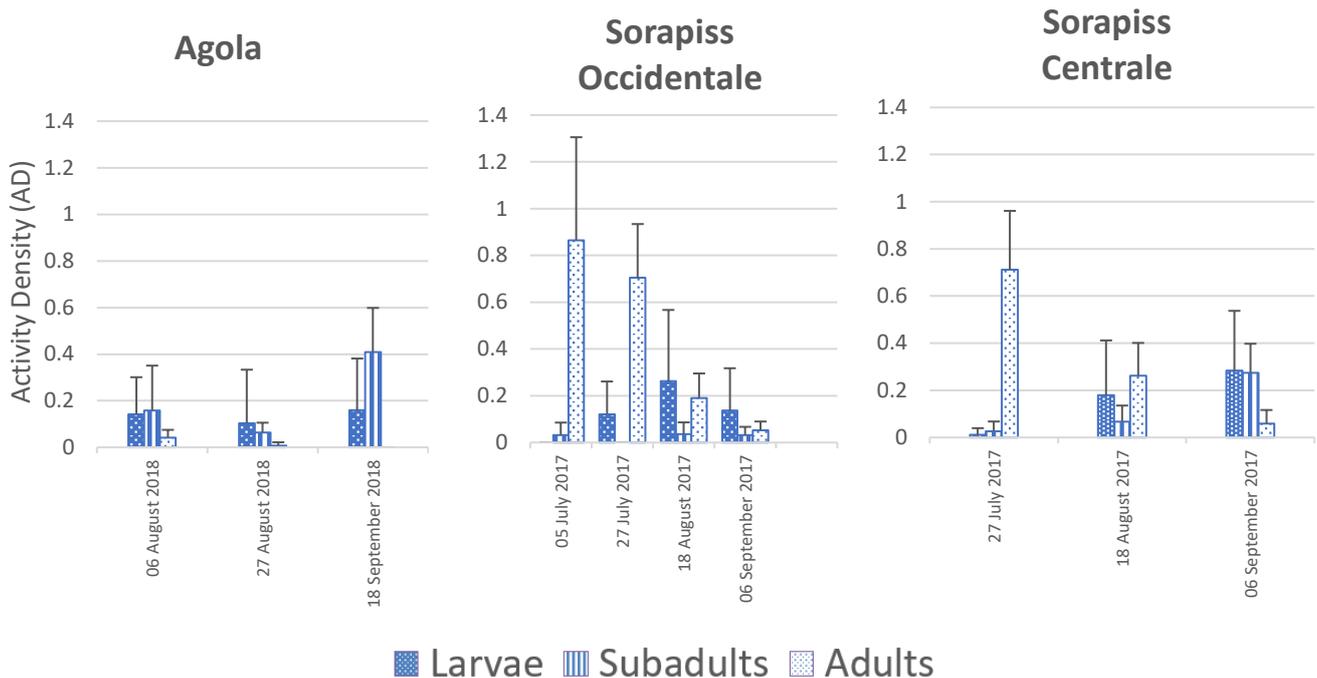

*Figure 3 Sampling data are expressed as average Activity Density (AD: number of individuals per day of trap activity). Whiskers represent standard deviation.*

# DISCUSSION

Data collected on the investigated glaciers of the Dolomites provided the first evidence about the micrometeorological features of the supraglacial habitats inhabited by *N. germari*. In addition, they gave us the opportunity to describe population patterns in time and in relation to potential competitors and to prey availability.

- MICROMETEOROLOGICAL CONDITIONS

Micrometeorological data gave detailed information about the cold conditions characterising supraglacial debris where *N. germari* currently lives. The recorded annual average temperatures always spanned between -1°C and 0°C. These data is in agreement with those recorded on the debris cover of Amola Glacier (Presanella Group, Central-Eastern Alps; GOBBI *et al* 2017) suggesting similar thermal properties notwithstanding different lithology (tonalite, an igneous intrusive rock) of the debris of this latter glacier. Average temperatures recorded in the supraglacial habitat were similar or slightly lower than those on the adjacent ice-free debris, but always higher than those estimated for the atmosphere during winter, as an effect of the snow cover, and lower than both them during summer, probably as a direct effect of the presence of ice. Temperatures were almost identical during day and night, but this also occurred on the glacier forelands. Thus, temperature homogeneity between day and night is not due to the presence of ice, since it is not an exclusive feature of supraglacial environment, but probably to the thermal inertia of stony debris. Other important features recorded on supraglacial debris are the low annual temperature excursions and the quite constant high level of humidity, an important condition for a species like *N. germari* that lives where soil is drenched of water, as observed by BRANDMAYR and ZETTO BRANDMAYR (1988). Thus, the micrometeorological condition in the extreme environment of the debris cover of glacier are more stable than those of the close ice-free debris, and therefore probably more suitable for this species.

Literature data reported that *N. germari* lives only in drenched soils, but little is known about the influence of temperature on its activity. Indeed, a study performed on the Rocky Mountains by SLAYTER and SCHOVILLE (2016) demonstrated that cold and heat tolerances vary little among Nebria species along an elevation gradient and these variations are not generally associated with species turnover; in particular, heat tolerance of adults spans between 32.2°C and 37.0°C. Even if our dataloggers registered maximum hourly peak of 16°C on Agola, 10.2°C on Western Sorapiss and 6.9°C on Central Sorapiss, GOBBI *et al* (2017) found that temperatures in other sites where *N. germari* lives (Amola glacier) can reach up to 30°C. This information suggests that N. germari adults might survive at higher temperatures like those found by SLATYER and SCHOVILLE (2016). As suggested by HÅGVAR *et al* (2017) for Norwegian alpine ground beetles, the ecology of larvae and their temperature preferences could be the key element determining alpine Nebria survival and habitat choice (THIELE 1972). Personal observations on Alpine glaciers highlighted that occurrence of N. germari larvae is higher at the ice-debris interface, where debris pressure melts the ice, which is the place where their prey (springtails) occur at high density. Therefore, we hypothesised that larvae may represent the stage of Nebria development that is more linked to the presence of ice. Low temperatures of supraglacial debris could influence also biotic factors related to interspecific interactions (SLATYER & SCHOVILLE 2016), for example reducing the predation pressure on larvae due to the absence of other predators, as proposed by CURRIE *et al* (1996). However, heat tolerance of *N. germari*'s adults and larvae still needs to be assessed in laboratory to confirm these hypotheses.

- PHENOLOGY

Our results show that *N. germari* adults are abundant at the beginning of the snow-free period, and then progressively decrease during summer. Larvae found during the snow-melt period (early July) develop during summer (July-September), resulting in the increase of subadults observed in late season (August-September). During our first sampling session on Agola Glacier, we collected a low number of adults and proportionally more larvae than on the other glaciers, suggesting that on that glacier, larvae development had started already; the lack of early adults, which are the most abundant developmental stage on Sorapiss Glaciers, may explain the generally lower numbers on Agola Glacier. Our data suggest that *N. germari* can be considered a spring breeder: the reproduction seems to start at the very beginning of the snow-free period (end of June - early July) and larvae develop during the following months, from July to September at least (the date of our last sampling). These observations differ from those of BRANDMAYR & ZETTO BRANDMAYR (1988), who reported that *N. germari* was an autumn breeder, at least for the populations found in high alpine carbonate discontinuous grasslands, but it is consistent with KAUFMANN and JUEN's (2001) observations on a pioneer habitat on an Austrian glacier. Previous studies therefore support evidence of differences in breeding time on alpine grasslands with respect to glacier forelands and glaciers surface. We stress, however, that species inhabiting alpine habitats usually reproduce during the short now-free period, therefore the dichotomy between spring and autumn breeder may have little ecological relevance, since breeding period only spam from July to September (SOTA 1996). Our dataloggers allowed estimating a very short snow-free period (121 days on Agola, 130 days on Western Sorapiss, 105 days on Central Sorapiss). A long persistence of the snow cover is a constant of *N germari*'s habitat (BRANDMAYR & ZETTO BRANDMAYR 1988, GEREBEN, 1995). Overwintering larvae and a biennial life-cycle are key adaptive features to high alpine environments characterised by a very short snow-free season (SOTA 1996, ANDERSEN 1984). Both these features are already known for *N. germari* (DE ZORDO 1979, BRANDMAYR & ZETTO BRANDMAYR 1988, KAUFMANN & JUEN 2001) and are consistent with our observations. Indeed, both the large number of larvae and subadults found at the end of the snow-free season and the abundance of adults immediately after snowmelt (reported also by GEREBEN, 1995) suggest that also supraglacial populations have overwintering larvae and probably a biennial life-cycle. However, it is still unknown if *N. germari* larvae enter diapause or remain active during winter. Indeed, even winter could potentially be a favourable growing season for N. germari, because environmental conditions are stable under the snow cover, and food may be available, due to cold-hardiness of Collembola. These features may allow either winter activity or dormancy (VANIN & TURCHETTO 2007, BLOCK & ZETTEL 1980, SØMME 1981, WAUDE & VERHOEF 1988). Which overwintering strategy is actually used by *N. germari* should therefore be tested in laboratory conditions

- RESPONSES TO ECOLOGICAL FACTORS

GLM analysis evidences that different life-stages respond to different predictors; this is reasonable if we consider that larvae have lower mobility (HÅGVAR *et al* 2017) and a different vertical distribution than adult and subadults (KAUFMANN & JUEN 2001). In addition, larvae and subadults may be more vulnerable to predation than adults because of the less sclerotized exoskeleton.

Adults density on the supraglacial debris was significantly related to grain size distribution; in particular, the gravel fraction was negatively related to *N. germari* adult abundance probably because of a lower detection probability where grain size is greater as suggested by TENAN *et al* (2016).

Subadult abundance was negatively related to spider abundance, maybe because of intraguild predation of spiders on subadult ground beetles. Differently from larvae that live deep in the debris, and the adults that have thicker exoskeleton, sub-adults have a soft exoskeleton and inhabit the surface, probably experiencing a higher predatory pressure. Alternatively, this association may be spurious and related to independent temporal trends of *N. germari* subadults and spiders. The very similar temperatures during day and night allow hypothesizing that *N. germari* could be active even during the day in the supraglacial habitat, thus being huntable by the spiders. Abundance of larvae was positively related to that of Centipedes even if they are both predators, probably because they co-occur in the same site where food is abundant. Indeed, they share similar edaphic needs and they probably live at the same depth. However, it should be considered that centipedes are very few (only 5 specimens on Agola and 4 on Central Sorapiss). In order to fully understand the interactions of *Nebria germari* with other predators, direct observations in the field and gut content analysis are therefore necessary (RASO *et al* 2014, SINT *et al* 2018).

Contrary to our expectations, we found no significant relation between *N. germari* density and springtail abundance. Springtails were often very abundant in our samples, particularly those belonging to the genus Orchesella that can represent an ideal prey for *N. germari*. We hypothesised that the lack of relation could be due to a general large abundance of surface-active springtails on the supraglacial debris, which therefore do not constitute a limiting factor for *N. germari*.

- CHANGES IN ALTITUDINAL DISTRIBUTION

*N. germari* seems to have moved to higher altitude, probably as a consequence of to climate change. Indeed, previous works (MARCUZZI 1956, BRANDMAYR & ZETTO BRANDMAYR 1988, PIZZOLOTTO *et al* 2014) reported that, until the second half of the last century, the lower altitudinal limit of *N. germari* in Dolomites was around 2000 m asl. Specifically, the species was found in discontinuous Alpine grasslands on carbonate substrata (dominated by Carex firma). Field data collected in the last thirty years on the Paneveggio-Pale di San Martino Dolomites (PIZZOLOTTO *et al* 2014) demonstrated the current extinction in this grassland type of the Southern Alps and an altitudinal shift of 300 meters in 28 years (1980-2008/09), with a consequent contraction of the distribution range. Our observations confirm the absence of this species from alpine grassland and the migration to more pioneer communities. In particular, we observed *N. germari* only on the supraglacial debris (excluding sporadic catches on the nearby glacier foreland; personal observations).

The minimum altitude in which this species occurs in the studied areas is currently 2598 m asl on Agola (Brenta Dolomites), 2264 m asl on Western Sorapiss, 2168 m asl on Central Sorapiss (Ampezzo Dolomites). In order to calculate the altitudinal shifts, we compared our data with historical work of MARCUZZI (1956) and with Marcuzzi's Collections preserved at Natural History Museum of Genova (10 specimens; sampling year: 1972) for Ampezzo Dolomites and with MUSE – Science Museum's collections (180 specimens; sampling years: from 1931 to 1958) for Brenta Dolomites. The average altitudinal shift occurred in c. 60 years was of c. 210 m in Ampezzo Dolomites (1956-2017; c. 260 m in Western Sorapiss and 170 m in Central Sorapiss) and of c. 600 m in Brenta Dolomites (1958 – 2018).

- ALPINE REFUGIA

In the last decades, the key biological role of debris-covered glaciers has been recognized. Indeed, this habitat is hosting cold-adapted arthropod species during the current interglacial period (HÅGVAR *et al* 2020; GOBBI *et al* 2018, GOBBI *et al* 2011). Dolomites are particularly sensitive to the effects of climate change because of their relatively low altitude and the small size of their glaciers, which are undergoing a shrinking of surface size and an increase in debris cover that could partially reduce melting rate (NAKAWO & RANA 1999). Here, *N. germari*, currently finds a suitable habitat only on the supraglacial debris, that persists at different altitudes depending on the glacier morphology and the degree of debris cover (PELFINI *et al* 2012): on Central Sorapiss the debris cover 70% of the glacier surface and protect the ice from melting even at an altitude as low as 2213 m asl. Here we observed the minimum altitudinal shift in *N. germari* distribution.

We found that glacier debris cover (of both active and inactive glaciers, like Central Sorapiss) has peculiar micrometeorological features that make this habitat a perfect refugium for a cold-adapted hygrophile species like *N. germari* during the current interglacial period. However, depending on the climatic regime, different landforms can assume the same role (TAMPUCCI 2017): in more continental and cold climates rock glaciers can persist at lower altitudes than debris-covered glacier, and can provide an ideal refuge also for subterranean species, because they are characterized by a thickness of the debris that can exceed 6 m with a considerable fissure network among boulders, which can include human-sized caves (GOBBI *et al* 2014).

The first satellite mapping of supraglacial debris of glacial and periglacial landform was provided by SCHERLER *et al* (2018) and allowed estimating that the extent of *N. germari* potential habitat in Eastern Alps is 555 km2, but extremely discontinuous and fragmented. This habitat is currently expanding, providing refuge to cold-adapted species. However, we do not know how long these landforms will last if the climate continues to warm. The risk of a complete loss of suitable habitats for this species is exacerbated by the fact that *N. germari* is present in Dolomites with the steno-endemic subspecies simony Ganglbauer, 1892; thus, its extinction in the Alps will cause an important loss for this "cold spot" of biodiversity.

## ACKNOWLEDGEMENTS

We thank Michael Bernasconi, Marina Serena Borgatti and Marta Tognetti for the key support during sampling sessions. We also thank Ampezzo Dolomites Natural Park and the Adamello-Brenta Natural Park for permission for carrying out this research.

# Appendix 1: sampling plan

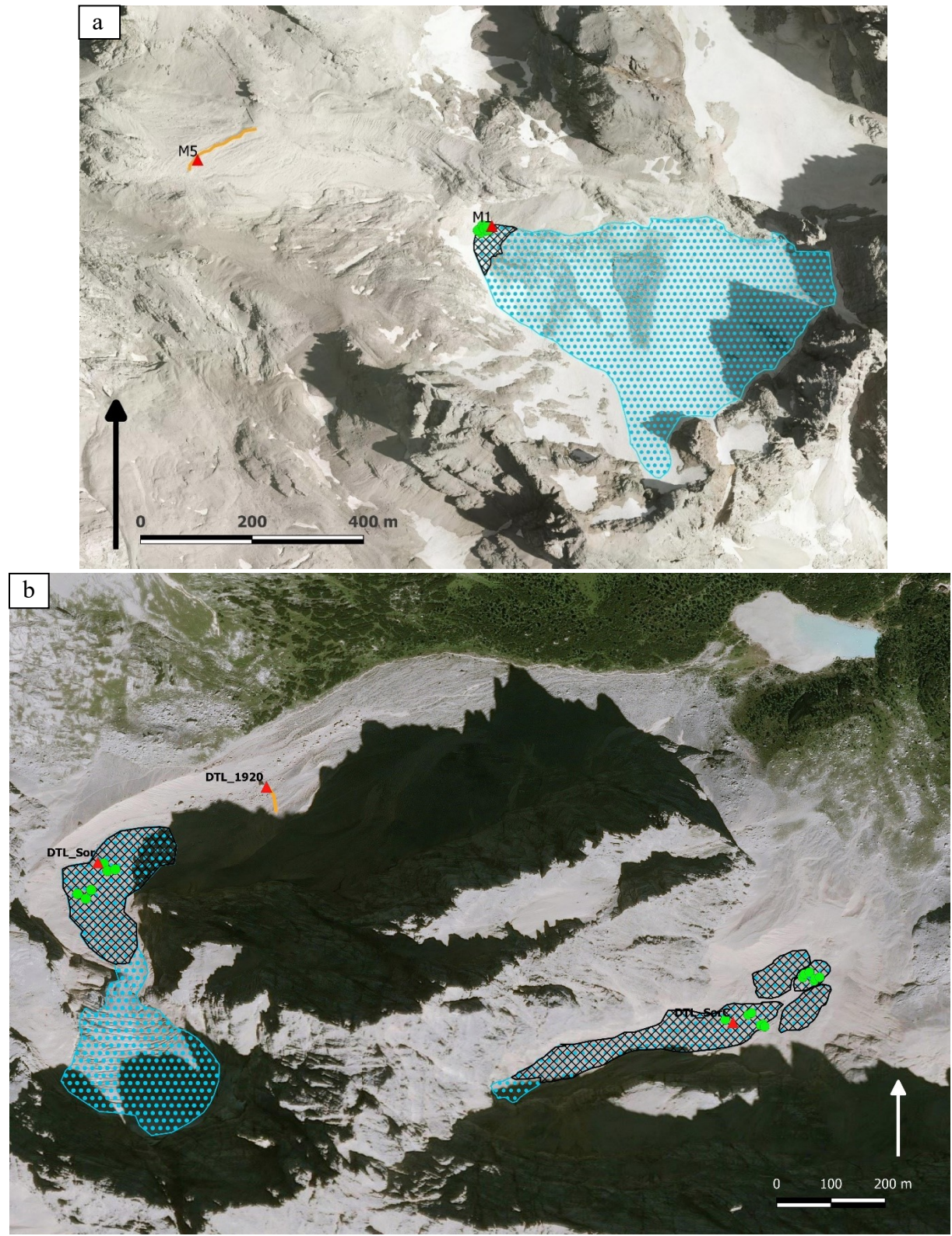

**Appendix 1:** sampling plan map of a) Agola glacier and b) Sorapiss glaciers. Light-blue pointed polygons: permanent ice; reticulated polygons: supraglacial debris; green points: pitfall traps; red triangles: Dataloggers, in partocular: M1 = supraglacial datalogger of Agola,

M5 = datalogger of 1908 Moraines of Agola, DTL_Sor = supraglacial datalogger of Sorapiss Occidentale, DTL_SorC = supraglacial datalogger of Sorapiss Centrale, DTL 1920 = datalogger of 1920 Moraines of Sorapiss.

# Appendix 2: average monthly temperatures

| | | \multicolumn{12}{c}{Average monthly temperature (°C)} | annual excursion (°C) | annual average (°C) |
|---|---|---|---|---|---|---|---|---|---|---|---|---|---|---|---|
| | | A (2018) | S (2018) | O (2018) | N (2018) | D (2018) | J (2019) | F (2019) | M (2019) | A (2019) | M (2019) | J (2019) | J (2019) | | |
| **AGOLA** Altitude: 2600 m | SUPRAGLACIAL DEBRIS | 4.8 | 2.8 | 0.3 | 0.0 | -0.6 | -1.7 | -2.4 | -2.2 | -1.0 | -0.2 | miss | miss | 7.2 | miss |
| | ATMOSPHERE | 9.0 | 5.9 | 2.2 | -4.3 | -4.1 | -7.1 | -1.3 | -1.6 | 0.0 | 1.3 | 10.9 | 10.7 | 18.0 | 1.8 |
| | ICE-FREE DEBRIS | 10.6 | 8.6 | 4.2 | 0.5 | 0.3 | -0.1 | -0.3 | -0.2 | -0.2 | -0.2 | -0.1 | 9.2 | 10.9 | 2.7 |
| | | J (2017) | A (2017) | S (2017) | O (2017) | N (2017) | D (2017) | J (2018) | F (2018) | M (2018) | A (2018) | M (2018) | J (2018) | | |
| **WESTERN SORAPISS** Altitude: 2360 | SUPRAGLACIAL DEBRIS | 5 | 5.1 | 0.8 | 0.5 | -1.4 | -2.6 | -2.6 | -2.8 | -2.9 | -1.6 | 0 | 0 | | 8.0 | -0.3 |
| | ATMOSPHERE | 6.6 | 8.0 | 1.3 | 1.4 | -5.4 | -9.2 | -8.4 | -12.3 | -7.6 | -0.7 | 1.7 | 4.8 | | 20.3 | -1.7 |
| | ICE-FREE DEBRIS | 8.1 | 9.3 | 2.8 | 1.5 | -1.8 | -3.1 | -3.1 | -3.2 | -2.7 | -1.8 | -0.1 | 5 | | 12.4 | 1.0 |
| | | | A (2017) | S (2017) | O (2017) | N (2017) | D (2017) | J (2018) | F (2018) | M (2018) | A (2018) | M (2018) | J (2018) | J (2018) | | |
| **CENTRAL SORAPISS** Altitude: 2360 | SUPRAGLACIAL DEBRIS | | 3 | 0.2 | -0.2 | -1.2 | -2.3 | -2.3 | -2.3 | -2.4 | -1.6 | 0 | 0 | 0.3 | 5.4 | -0.8 |
| | ATMOSPHERE | | 8.5 | 1.8 | 1.8 | -5.0 | -8.9 | -6.9 | -11.8 | -7.3 | -0.1 | 2.3 | 5.3 | 7.7 | 20.3 | -1.0 |
| | ICE-FREE DEBRIS | | 9.8 | 3.3 | 1.9 | -1.4 | -2.8 | -1.4 | -2.8 | -2.4 | -1.2 | 0.5 | 5.6 | 8.6 | 12.6 | 1.5 |

Appendix 2: monthly temperatures, annual average temperature and annual excursion on supraglacial debris, ice-free debris and atmosphere of Agola, Central and Western Sorapiss Glaciers. Monthly temperature is the average of daily temperatures per month.

# Appendix 3: daily temperature values

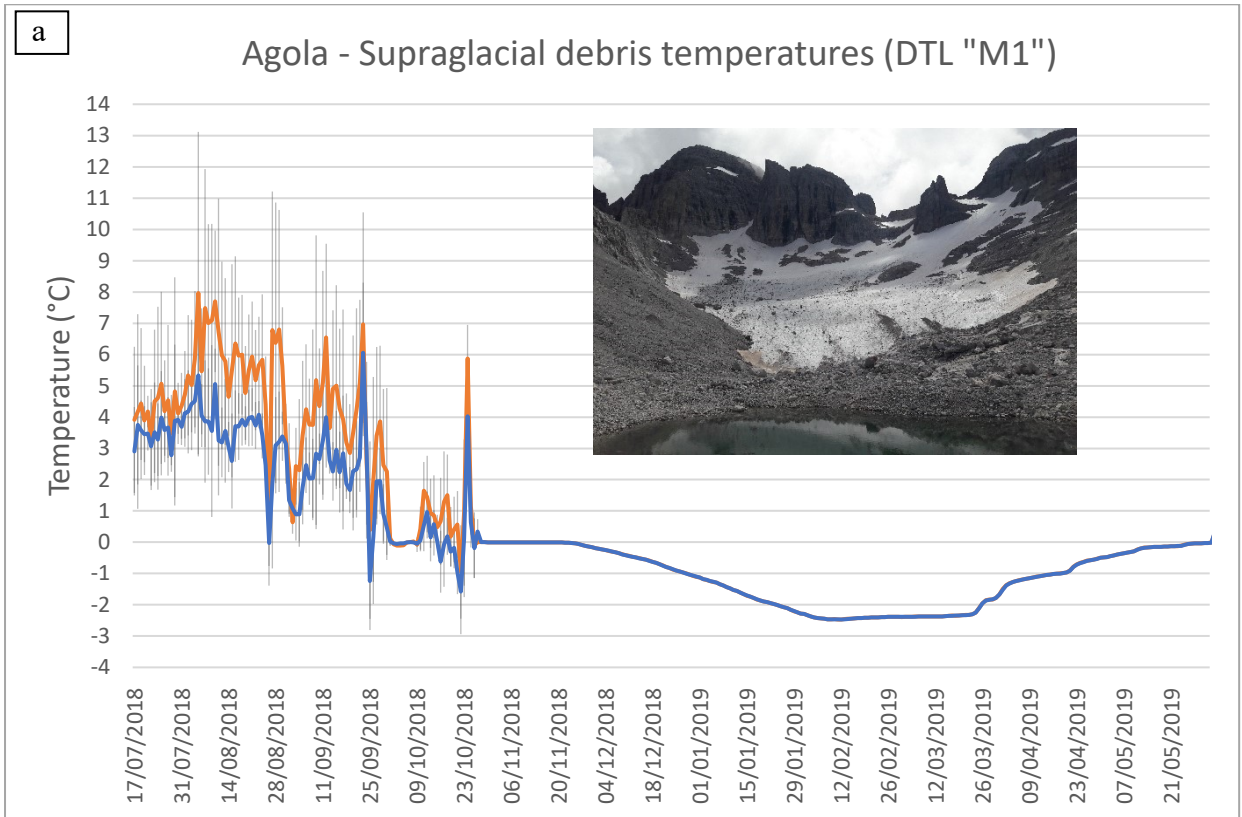

a) Agola - Supraglacial debris temperatures (DTL "M1")

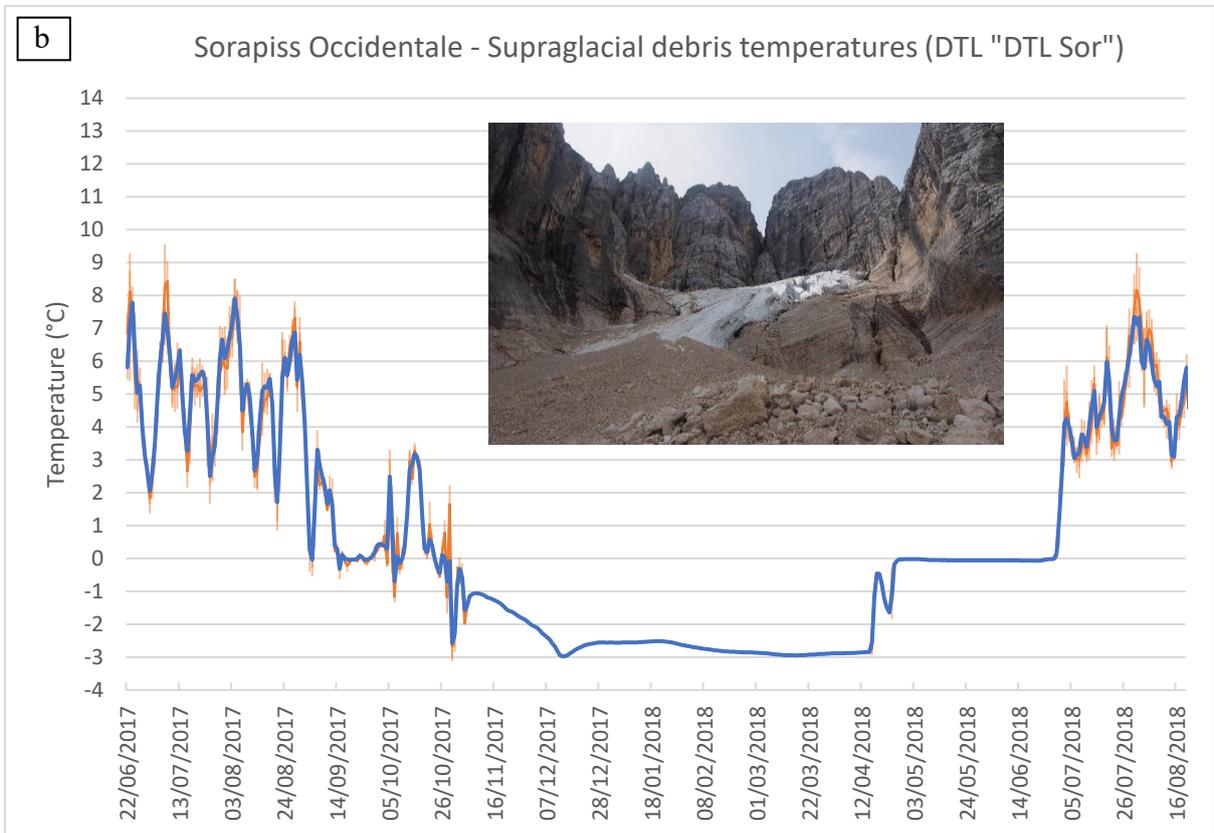

b) Sorapiss Occidentale - Supraglacial debris temperatures (DTL "DTL Sor")

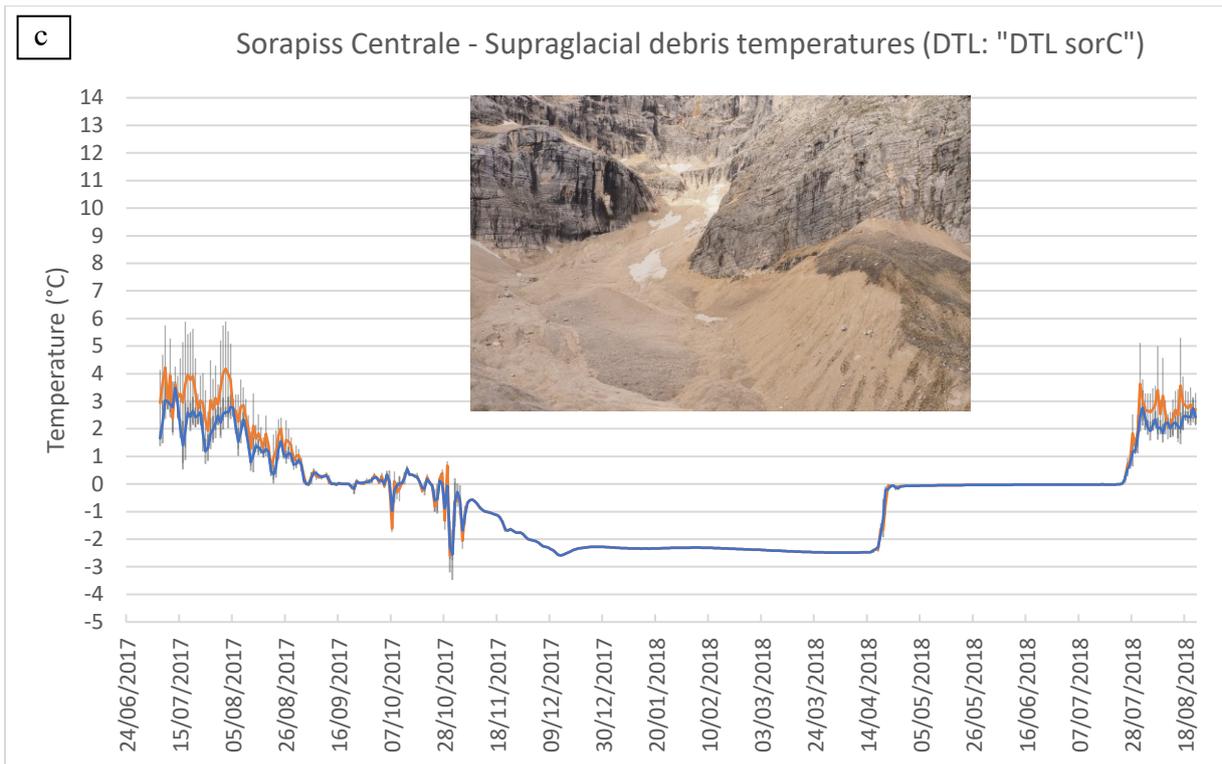

Appendix 3: graphs report temperature values recorded on supraglacial debris of a) Agola, b) Sorapiss Occidentale and c) Sorapiss Centrale. Orange lines indicate average day values, blue lines indicate average night values. Grey bars reported standard deviation of daily values.